\documentclass[12pt]{article}
\usepackage{amsmath, amssymb, theorem}
\usepackage{graphicx}
\usepackage{subfigure}
\usepackage{epstopdf}
\usepackage[font={small,it}]{caption}

\newcommand{\VAR}{\mathrm{VaR}}
\newcommand{\ES}{\mathrm{ES}}

\newcommand{\cL}{\mathcal{L}}

\newcommand{\eps}{\varepsilon}
\renewcommand{\leq}{\leqslant}
\renewcommand{\geq}{\geqslant}

\newcommand{\mR}{\mathbb{R}}
\newcommand{\infrho}{\underline{\rho}}
\newcommand{\suprho}{\overline{\rho}}
\newcommand{\AMR}{\mathrm{AM}}
\newcommand{\RMR}{\mathrm{RM}}

\newcommand{\proof}[1]{\noindent\textbf{Proof.
}#1\hfill$\Box$\\\par}
\theoremstyle{plain}
\theorembodyfont{\it}
\newtheorem{theorem}{theorem}[section]
\newtheorem{corollary}[theorem]{Corollary}
\newtheorem{lemma}[theorem]{Lemma}
\newtheorem{proposition}[theorem]{Proposition}
\theorembodyfont{\rm}
\newtheorem{definition}[theorem]{Definition}
\newtheorem{remark}[theorem]{Remark}

\begin{document}
\title{Assessing Financial Model Risk}
\author{Pauline Barrieu\footnote{Department of Statistics, London School of Economics (\texttt{p.m.barrieu@lse.ac.uk})} \qquad Giacomo Scandolo\footnote{Department of Economics, University of Verona (\texttt{giacomo.scandolo@univr.it})}}
\date{June 24, 2013}
\maketitle
\abstract{Model risk has a huge impact on any risk measurement procedure and its quantification is therefore a crucial step. In this paper, we introduce three quantitative measures of model risk when choosing a particular reference model within a given class: the absolute measure of model risk, the relative measure of model risk and the local measure of model risk. Each of the measures has a specific purpose and so allows for flexibility. We illustrate the various notions by studying some relevant examples, so as to emphasize the practicability and tractability of our approach.}

\section{Introduction}
The specification of a model is a crucial step when measuring financial risks to which a portfolio is exposed. Common methodologies, such as Delta-Normal or simulation methods, are based on the choice of a particular model for the risk factors. Even when using historical methods, we implicitly rely on the empirical distribution as the reference model. However, it is observed that the final risk figure is often quite sensitive to the choice of the model. The hazard of working with a potentially not well-suited model is referred to as \emph{model risk}. The study of the impact of model risk and its quantification is an important step in the whole risk measurement procedure. In particular, in the aftermath of the recent financial crisis, understanding model uncertainty when assessing the regulatory capital requirements for financial institutions seems to be crucial. The main goal of this paper is precisely to propose some ways to quantify model risk when measuring financial risks for regulatory purposes. We stress that our objective is \emph{not} to measure risk in the presence of model uncertainty, but to quantify model risk itself.

The question of the impact of model risk has received increasing attention in recent years. In particular, the significance of minimum risk portfolios has been questioned when studying the problem of optimal asset allocation: several authors (among them El Ghaoui et al. 2003, Natarajan et al. 2008, Chen et al. 2010, Zymler et al. 2013) have recently considered this issue from a robust optimization perspective.

Our approach to assessing model risk is very general. It is based on the specification of a set of alternative models (or distributions) around a reference one. Note that Kerkhof et al. (2010) propose measuring model risk in a similar setting by computing the worst-case risk measure over a \emph{tolerance set} of models. Our approach differs, however, as we introduce different measures of model risk, based on both the worst- and best-case risk measures, in order to serve different purposes.

Examples of the set of alternative models we can consider include parametric or non-parametric families of distributions, or small perturbations of a given distribution. If we believe in a parametric model, we can consider all distributions within the family whose parameters are in the confidence intervals derived from the data. By doing this, we are accounting only for the \emph{estimation risk} (see Kerkhof et al. 2010). If, on the other hand, we completely believe in some estimated quantities (for instance, mean and variance), without relying on confidence intervals, we can consider all possible distributions of any form which are in accordance with those quantities (for instance, they have the same mean and variance). We can also consider those distributions which are not too far from a reference one, according to some statistical distance (the uniform distance, for instance), or all joint distributions that have the same marginals as the reference one. This latter example leads to the relevant problem of aggregation of risks in a portfolio (see Embrechts et al 2013). We could even specify different pricing models if the portfolio contains derivatives.

Note that the scope of our approach is very wide, going beyond issues pertaining just to statistical estimation. Furthermore, the assessment of model risk should not be confused with the analysis of statistical robustness of a risk measurement procedure (as in Cont et al. 2010), even though the two concepts are related. Indeed, the reference distribution is an input in our approach, while in Cont et al. (2010) it is the result of a statistical estimation process which is part of the definition of robustness itself.

In order to assess model risk, we introduce three different measures: the absolute measure of model risk, the relative measure of model risk and the local measure of model risk. Our aim is to provide a quantitative measure of the model risk we are exposed to in choosing a particular reference model within a given class when working with a specific risk measure. All three measures are pure numbers, independent from the reference currency. They take non-negative values and vanish precisely when there is no model risk. Each of the measures we propose has a specific purpose: whilst the absolute measure is cardinal and gives a quantitative assessment of model risk, both the relative measure and the local measure are ordinal and allow for comparison of different situations, which may have different scales. If we consider different possible models as references, the use of the relative measure is probably the more natural measure to use as it will give a clear ranking between the alternatives. When the reference model is almost certain, the local measure becomes an obvious choice as it focuses on the very local properties around the reference model.

In addition, we obtain explicit and closed-form formulae in some interesting situations when considering either the Value-at-Risk or the Expected Shortfall as reference risk measure and alternative sets of distributions based on fixed moments or small perturbations based on some standard statistical distances.

\section{A motivating example}
In this section, we start by looking at the Basel multiplier, introduced by the Basel Committee as an ingredient in the assessment of the capital requirements for financial institutions. As we will see, this multiplier is closely related to probabilistic bounds giving some upper limit to classical risk measures such as the Value-at-Risk and the Expected Shortfall. These preliminary remarks will motivate our approach when introducing some measures for model risk in the next section.

\subsection{The Basel multiplier}
Within the Basel framework, financial institutions are allowed to use
internal models to assess the capital requirement due to market risk. The
capital charge is actually the sum of six terms taking into account
different facets of market risk. The term that measures risk in \emph{usual}
conditions is given by the following formula:
\begin{equation}  \label{charge}
CC = \max\left\{\mathrm{VaR}^{(0)},\frac{\lambda}{60}\sum_{i=1}^{60}\mathrm{VaR}^{(-i)}\right\},
\end{equation}
where $\mathrm{VaR}^{(0)}$ is the portfolio's Value-at-Risk (of order $1\%$
and with a $10$-day horizon) computed today, while $\mathrm{VaR}^{(-i)}$ is
the figure we obtained $i$ days ago.

The constant $\lambda$ is called the
\emph{multiplier} and it is assigned to each institution by the regulator,
which periodically revises it. Its minimum value is $3$, but it can be
increased up to $4$ in the event that the risk measurement system provides poor
back-testing performances. Given the magnitude of $\lambda$, it is apparent that in normal conditions
the second term is the leading one in the maximum appearing in \eqref{charge}.
\subsection{Chebishev bounds and the multiplier}
Stahl (1997) offered a simple theoretical justification for the multiplier
to be chosen in the range $[3,4]$. Here, we briefly summarize his argument.
Let $X$ be the random variable (r.v.) describing the Profits-and-Losses of a
portfolio due to market risk. If the time-horizon is short, it is usually
assumed that $\mathbb{E}[X]=0$, so that
\begin{equation*}
\mathrm{VaR}_\alpha(X)=\sigma\mathrm{VaR}_\alpha(\widetilde{X}),
\end{equation*}
where $\sigma^2$ is the variance of $X$ and $\widetilde{X}=X/\sigma$ is
\emph{standard}, i.e. it has zero mean and unit variance. While $\sigma$ is
a matter of estimation, $\mathrm{VaR}_\alpha(\widetilde{X})$ depends on the
assumption we make about the \emph{type} of the distribution of $X$ (normal,
Student-t, etc.).

\begin{figure}[htbp]
 \centering
 \label{cap1-fig1}
 \subfigure[$\VAR_\alpha$]
   {\includegraphics[width=7.8cm,height=7cm]{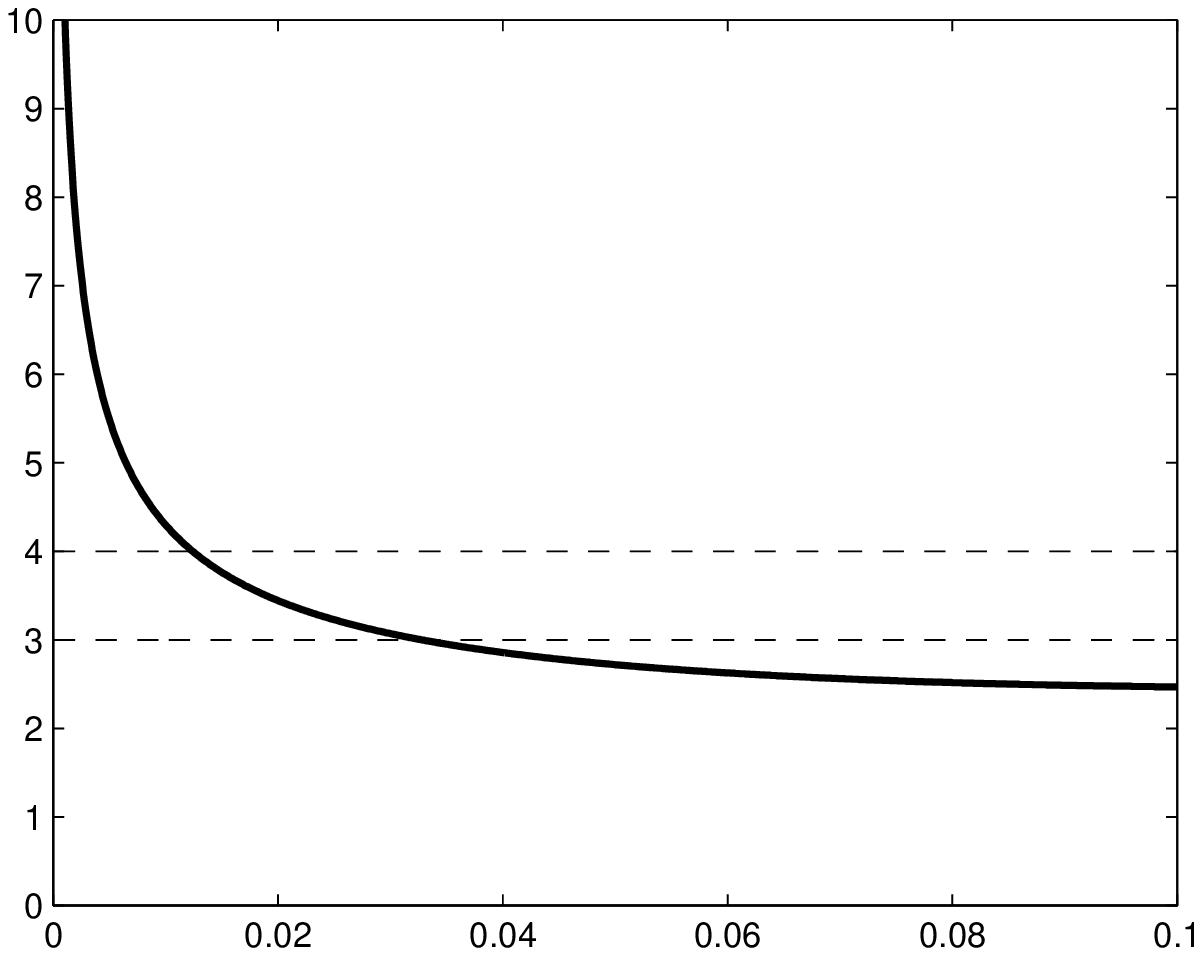}}
 \subfigure[$\ES_\alpha$]
   {\includegraphics[width=7.8cm,height=7cm]{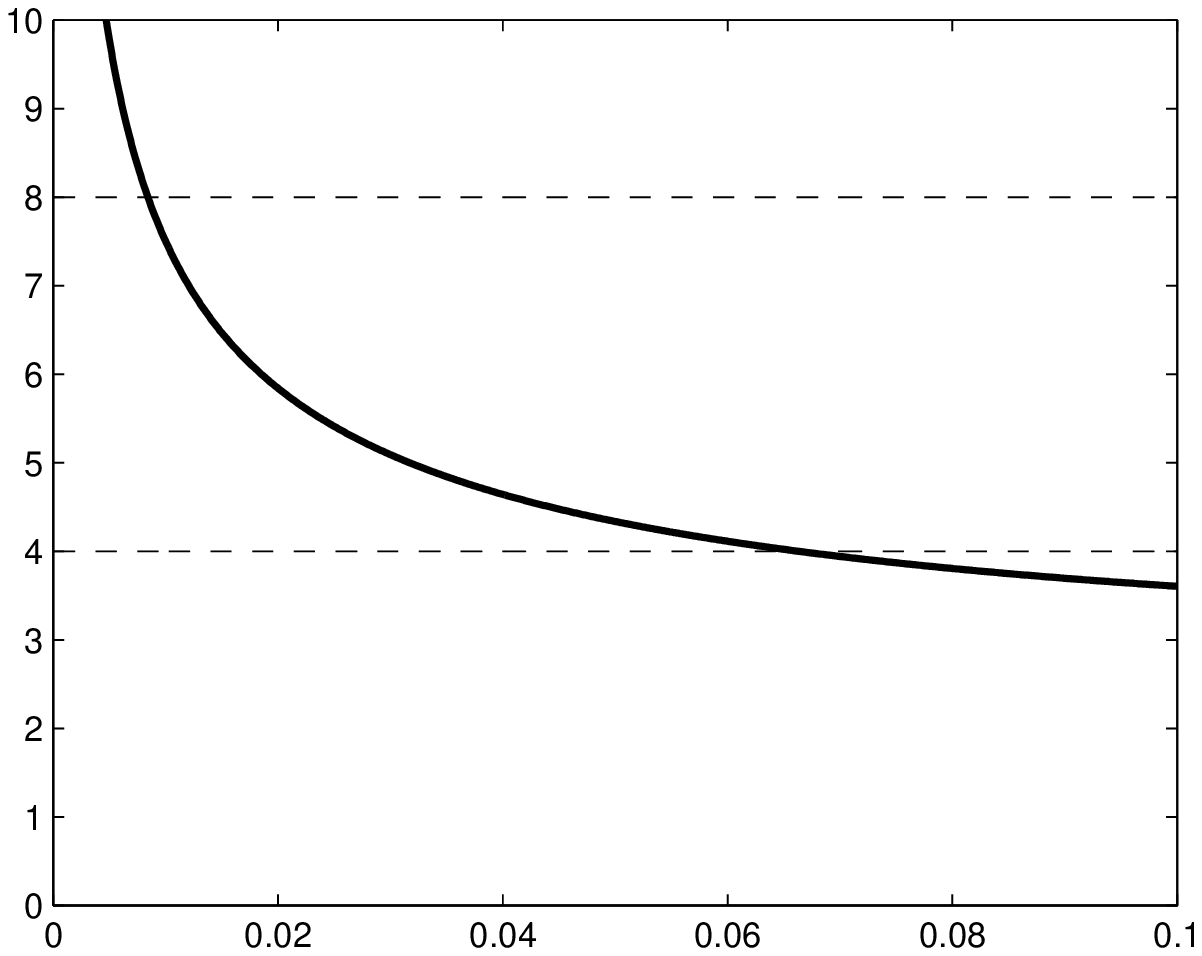}}
 \caption{Ratio, as a function of $\alpha\in(0,10\%)$, between the upper Chebishev bound and the risk measure under Gaussian hypothesis. \label{figAB}}
 \end{figure}

An application of the Chebishev inequality to $\widetilde{X}$ yields
\begin{equation}  \label{cheb}
P(\widetilde{X}\leq -q)\leq P(|\widetilde{X}|\geq q)\leq \frac{1}{q^2},\quad
q>0.
\end{equation}
Recalling the definition of $\mathrm{VaR}$, it readily follows $\mathrm{VaR}_\alpha(\widetilde{X})\leq1/\sqrt{\alpha}$, or
\begin{equation}  \label{ineq-var}
\mathrm{VaR}_\alpha(X)\leq \frac{\sigma}{\sqrt{\alpha}}.
\end{equation}
The right hand side of the above inequality thus provides an upper bound for
the VaR of a random variable having mean $0$ and variance $\sigma^2
$. It can be compared with the VaR we obtain by using the delta-normal
method, which is very commonly employed in practice. According to this
method, $\widetilde{X}$ is normally distributed and therefore
\begin{equation*}
\mathrm{VaR}_\alpha(X)=\sigma|z_\alpha|\qquad(\alpha<0.5),
\end{equation*}
where $z_\alpha=\Phi^{-1}(\alpha)$ is the quantile of a standard normal. The
graph of the ratio
\begin{equation}  \label{ratio1}
\frac{\sigma/\sqrt{\alpha}}{\sigma|z_\alpha|}=\frac{1}{|z_\alpha|\sqrt{\alpha}}
\end{equation}
is reported below (see Figure \ref{figAB}, left). We can see that for usual
values of $\alpha$ (i.e. from $1\%$ to $5\%$), the ratio broadly lies in the
interval $[3,4]$. Therefore, if the VaR computed under normal assumptions is
multiplied by $\lambda$, we obtain an upper bound for the worst possible VaR compatible with partial information (mean and variance) we have.

We can then extend this argument to the Expected Shortfall.\footnote{Also see Leippold and Vanini (2002)}
Indeed, by integrating inequality \eqref{ineq-var}, we obtain
\begin{equation}  \label{ineq-es}
\mathrm{ES}_\alpha(X)=\frac{1}{\alpha}\int_0^\alpha \mathrm{VaR}_u(X)\,du\leq
\frac{\sigma}{\alpha}\int_0^\alpha \frac{du}{\sqrt{u}}=\frac{2\sigma}{\sqrt{\alpha}}.
\end{equation}
The upper bound has to be compared with the Expected Shortfall under normal
assumptions, which is
\begin{equation*}
\mathrm{ES}(X)=\frac{\sigma\varphi(z_\alpha)}{\alpha},
\end{equation*}
where $\varphi$ is the density of a standard normal. From the graph of the
ratio
\begin{equation*}
\frac{2\sigma/\sqrt{\alpha}}{\sigma\varphi(z_\alpha)/\alpha}=\frac{2\sqrt{\alpha}}{\varphi(z_\alpha)}
\end{equation*}
(see Figure \ref{figAB}, right) we see that a proper multiplier for the Expected Shortfall
would be in the range $[4,8]$.

The second inequality in \eqref{cheb} is
sharp, i.e. it cannot be improved for any $q$. However, the first inequality
is certainly not sharp and this means that the upper bounds for VaR and Expected Shortfall that
we derived above are not optimal ones.

\subsection{Cantelli bounds and improvement of the multiplier}\label{Subsection sharp bounds}
Better results for the bounds can be achieved by using the Cantelli inequality which
concentrates on a single tail. A possible version of this inequality states
that for a standard r.v. $\widetilde{X}$, the following inequality holds true:
\begin{equation}  \label{cantelli}
P(\widetilde{X}\leq -q)\leq \frac{1}{1+q^2},\qquad q>0.
\end{equation}
From \eqref{cantelli} it readily follows that
\begin{equation}  \label{ineq-var2}
\mathrm{VaR}_\alpha(X)\leq\sigma\sqrt{\frac{1-\alpha}{\alpha}}
\end{equation}
for any random variable having mean $0$ and variance $\sigma^2$. We see that
this latter bound improves on \eqref{ineq-var}. Nevertheless, the ratio
between this bound and the VaR computed under normal assumptions broadly
remains between $3$ and $4$.

Integrating \eqref{ineq-var2} we obtain the following upper bound
for the Expected Shortfall:
\begin{equation}\label{ES-cantelli}
\mathrm{ES}_\alpha(X)\leq\frac{\sigma}{\alpha}\int_0^\alpha\sqrt{\frac{1-u}{u}}\,du= \frac{\sigma}{\alpha}\left(\sqrt{\alpha-\alpha^2}+\arctan\sqrt{\frac{1-\alpha}{\alpha}}\right).
\end{equation}
This bound slightly improves on \eqref{ineq-es}.
\subsection{Sharp bounds and significance of the multiplier}
It is well known that the Cantelli inequality provides a \emph{sharp} upper bound on the tail probability.\footnote{See for instance Billingsley (1995), Section 5.} To put it another way, the following holds true:
\begin{equation*}
    \sup_{\widetilde{X} \text{ standard}}P(\widetilde{X}\leq -q)=\frac{1}{1+q^2},\qquad q>0.
\end{equation*}
This means that $\sqrt{(1-\alpha)/\alpha}$ is a sharp upper bound on $\VAR_\alpha(\widetilde{X})$ for $\widetilde{X}$ standard (see also Lemma \ref{Lemma extremal functions} below). By contrast, the bound \eqref{ES-cantelli}, being an integral of sharp bounds, is not necessarily sharp. Indeed, we will recall later that the sharp bound is, in this case, $\ES_\alpha(X)\leq \sqrt{(1-\alpha)/\alpha}$.

\begin{figure}[htbp]
 \centering
 \subfigure[$\VAR_\alpha$]
   {\includegraphics[width=7.8cm,height=7cm]{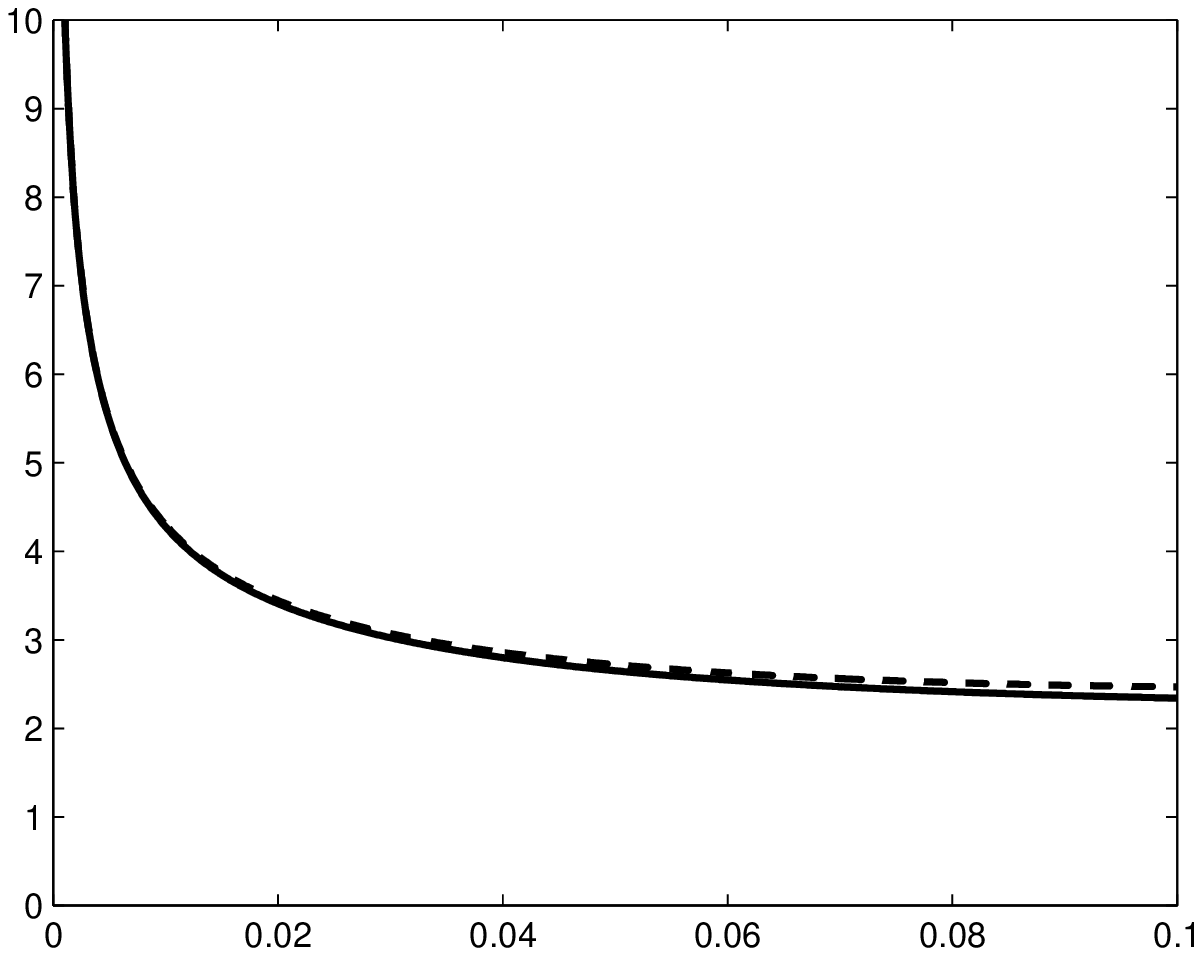}}
 \subfigure[$\ES_\alpha$]
   {\includegraphics[width=7.8cm,height=7cm]{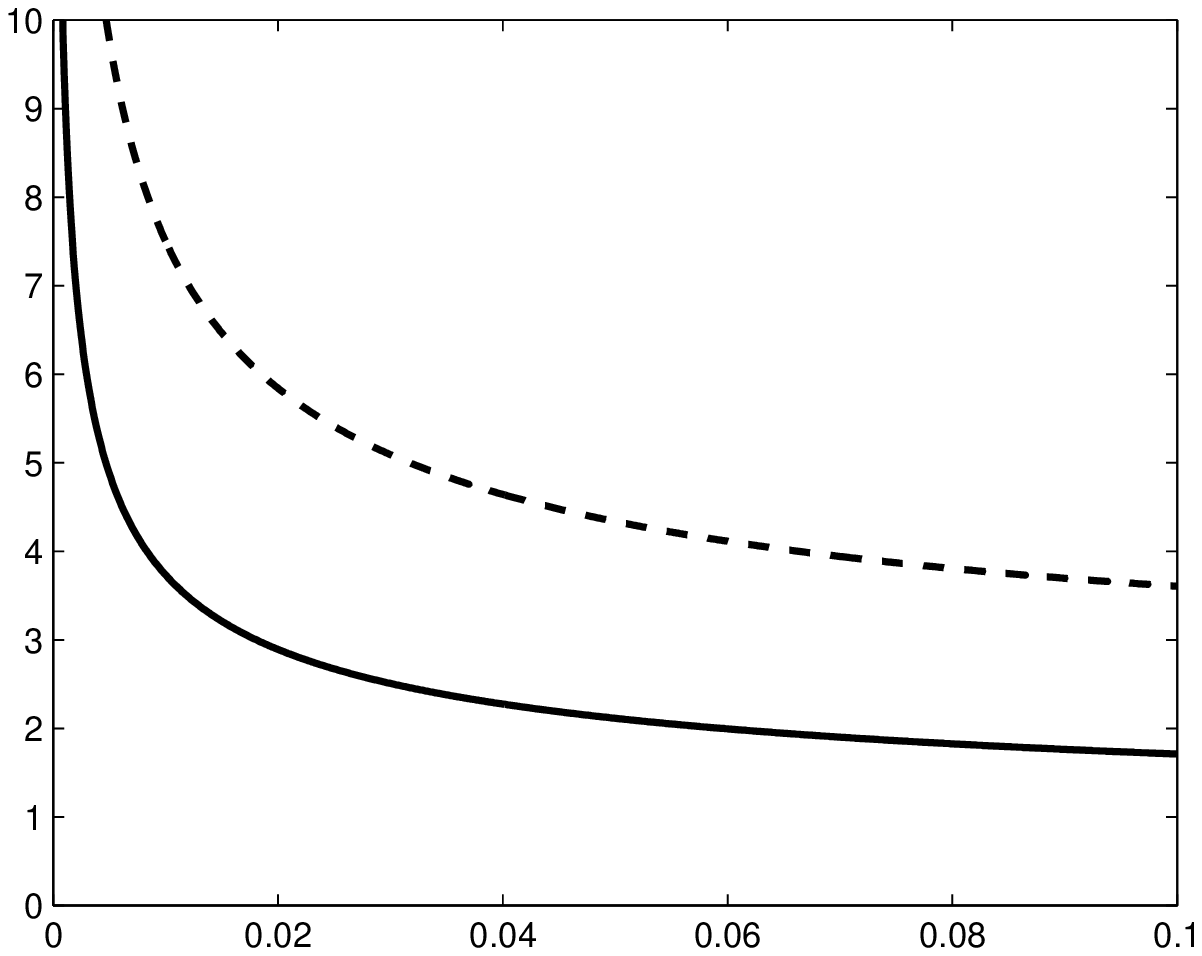}}
 \caption{Ratio between the Chebishev (dashed) and sharp (continuous) upper bound and the risk measure under Gaussian hypothesis. \label{figC}}
 \end{figure}
 %

We can plot the ratio between the sharp upper bound and the risk measure computed under Gaussian hypotheses and compare it with the ratio we obtained before, using the Chebishev bounds. The results are in Figure \ref{figC}.
We can notice that for the Expected Shortfall, the actual ratio (i.e. the one based on the sharp bound) is much lower than the ratio based on the Chebishev bound and the actual multiplier should be in the range $[3,4]$ for the Expected Shortfall as well. This also means that assessing the impact of model uncertainty using Chebishev bounds can give us misleading answers regarding the Expected Shortfall.

Therefore, it becomes apparent that an accurate analysis and understanding of the sharp bounds for the considered risk measure is essential in the assessment of model risk. Any other bounds may lead to an inaccurate assessment of the model risk and as a consequence to potential errors in any associated decision process. For that reason, in this paper we introduce different measures of model risk based on sharp bounds (both lower and upper bounds). The explicit computation of those bounds will then be a crucial step.
\section{Absolute and relative measures of model risk}
In this section, we introduce two different notions of measures of model risk. We will work with a given risk measure, a given reference model and a set of alternative models. Our aim is to provide a quantitative measure of the model risk we are exposed to in choosing this particular reference model within a given class when working with a specific risk measure. Two measures are introduced: the absolute measure of model risk provides a cardinal measure whilst the relative measure of model risk is ordinal and allows for comparison between various situations.
\subsection{Notation}
We first introduce some basic notation and assumptions to be used here and
in the sequel to this paper. A probability space $(\Omega,\mathcal{F},P)$ is
given and we assume it to be atomless.\footnote{This ensures, for any distribution $F$, the existence of a
r.v. distributed as $F$.} For any r.v. $X$ defined on $(\Omega,\mathcal{F},P)$, let $F_X$ be the associated distribution function, i.e. $F_X(x)=P(X\leq x)$, and
\begin{equation*}
q_\alpha(X)=\inf\{x\,:\,F_X(x)\geq \alpha\}
\end{equation*}
be the (lower) quantile of order $\alpha\in(0,1)$. We will write $X\sim Y$
if $F_X\equiv F_Y$ and $X\sim F$ if $F_X\equiv F$. In this paper, a \emph{risk measure} is a map $\rho:\cL_\rho\to\mathbb{R}$, defined on some
space of r.v. $\cL_\rho$ and satisfying the following properties
\begin{itemize}
\item \emph{law invariance}: $\rho(X)=\rho(Y)$ whenever $X\sim Y$

\item \emph{positive homogeneity}: $\rho(a X)=a\rho(X)$ for any $a\geq 0$

\item \emph{translation invariance}: $\rho(X+b)=\rho(X)-b$ for any $b\in\mathbb{R}$
\end{itemize}
We remark that, for fixed $\alpha\in(0,1)$, both the Value-at-Risk
\begin{equation*}
\mathrm{VaR}_\alpha(X)=-q_\alpha(X),
\end{equation*}
and the Expected Shortfall
\begin{equation*}
\mathrm{ES}_\alpha(X)=\frac{1}{\alpha}\int_0^\alpha \mathrm{VaR}_u(X)\,du
\end{equation*}
satisfy these assumptions. We stress that Value-at-Risk is defined over
\emph{all} random variables, while the Expected Shortfall requires an integrability condition on the left tail of $X$. More generally any law-invariant coherent risk measure falls
in our framework, a chief example being the class of spectral risk measures
(see Acerbi 2002). In view of the law invariance property, we can alternatively regard a risk measure as a functional directly defined on a suitable set of
distributions. Indeed, with a slight abuse of notation, we can set $\rho(F)=\rho(X)$ for $X\sim F$.
\subsection{Definitions}
We now introduce two measures of model risk. Both measures are associated to
a risk measure $\rho$, a r.v. $X_0$, to act as a reference
distribution hypothesis, and a set $\cL$ of r.v., to act
as alternative distribution hypotheses. In this paper, we do not discuss the selection procedure for the reference distribution, and refer to Alexander and Sarabia (2012), where some specific criteria are reviewed. We assume that $X_0\in \cL\subset \cL_\rho$. We also assume that both quantities
\begin{equation*}
\underline{\rho}(\cL)=\inf_{X\in \cL}\rho(X),\qquad
\overline{\rho}(\cL) =\sup_{X\in \cL}\rho(X)
\end{equation*}
are finite and that $\underline{\rho}(\cL)\neq \overline{\rho}(\cL)$. Clearly, the inequalities $\underline{\rho}(\cL)\leq
\rho(X_0)\leq \overline{\rho}(\cL)$ hold true. Finally, we assume that $\rho(X_0)>0$: this is not a restrictive hypothesis as the measured risk of
financial positions is usually positive. We are ready to give the two
definitions of model risk.
\begin{definition}
The \textbf{absolute measure of model risk} associated to $\rho$, $X_0$ and $\cL$ is\footnote{For the sake of simplicity, we drop the obvious dependence on $\rho$.}
\begin{equation*}
\mathrm{AM}=\mathrm{AM}(X_0,\cL)=\frac{\overline{\rho}(\cL)}{\rho(X_0)}-1.
\end{equation*}
The \textbf{relative measure of model risk} is
\begin{equation*}
\mathrm{RM}=\mathrm{RM}(X_0,\cL)=
\frac{\overline{\rho}(\cL)-\rho(X_0)}{\overline{\rho}(\cL)-\underline{\rho}(\cL)}.
\end{equation*}
\end{definition}
The absolute measure is a concept which in a sense generalizes the Basel
multiplier: indeed, by multiplying $\rho(X_0)$ by $\mathrm{AM} +1$ we reach
the maximum risk that is attainable within $\cL$. So, if we
interpret $\cL$ as a set of possible departures from the reference
model $X_0$, then $\mathrm{AM}$ quantifies how bad the worst possible case is. Plainly, $\mathrm{AM}\geq 0$ with $\mathrm{AM}=0$ (i.e. no model risk) if and only if $X_0$ has already a
worst-case distribution, i.e. $\rho(X_0)=\overline{\rho}(\cL)$.

It is apparent that, for given $\rho$ and $X_0$, the larger $\cL$
is the greater $\mathrm{AM}$ is, as $\overline{\rho}(\cL)$ is
increasing in $\cL$. This justifies the qualifier \emph{absolute}
that we give to $\mathrm{AM}$, even though it comes in the form of a ratio.

By contrast, $\mathrm{RM}$ has a \emph{relative} behaviour. Indeed, the
difference $\overline{\rho}(\cL)-\rho(X_0)$ is divided by the whole
range $\overline{\rho}(\cL)-\underline{\rho}(\cL)$. As a
consequence, it is immediately seen that
\begin{equation*}
0\leq \mathrm{RM}\leq 1.
\end{equation*}
We observe $\mathrm{RM}=0$ or $1$ precisely when $\rho(X_0)=\overline{\rho}(\cL)$ (no model risk) or $\rho(X_0)=\underline{\rho}(\cL)$
(full model risk). In other words, it focuses on the \emph{relative position}
of $\rho(X_0)$ within the range $[\underline{\rho}(\cL),\overline{\rho}(\cL)]$ and not only on the position with respect to the
supremum. In the next section, we will also see that $\mathrm{RM}$ need not
be increasing in $\cL$, thus providing a \emph{relative} assessment
of model risk.
\begin{remark}
Using the previous notation, the measure of model risk introduced in Kerkhof et al (2010) is
\begin{equation*}
   M_K=\overline{\rho}(\cL)-\rho(X_0).
\end{equation*}
We note that this measure is also non-negative and vanishes precisely when there is no model risk. However, it is expressed in terms of a given currency and depends on the scale of the risk $X_0$. Since $\mathrm{AM}=M_K/\rho(X_0)$, the absolute measure proposed here is a unit-less version of $M_K$, normalized by the size of the risk. We think that this normalization allows us to use $\mathrm{AM}$ also as a comparison tool between different situations.
\end{remark}
\begin{remark}
In the different context of derivative pricing, Cont (2006) proposed a measure of model risk which is based on the computation of extremal prices using a set of pricing measures. The obtained measure is formally similar to our definitions.
\end{remark}
\subsection{Properties}
In the next proposition, we collect some basic properties of the two measures
of model risk previously introduced. For any $a,b\in\mathbb{R}$ we define
\begin{equation*}
a\cL+b=\{aX+b\,:\,X\in \cL\}
\end{equation*}
\begin{proposition}
\label{properties-1} For any $a>0$ and $b\in\mathbb{R}$ it holds
\begin{align*}
\mathrm{AM}(aX_0,a\cL)&=\mathrm{AM}(X_0,\cL), \\[1mm]
\quad\mathrm{AM}(X_0+b,\cL+b)&\left\{
\begin{array}{ll}
>\mathrm{AM}(X_0,\cL), & \text{for }b>0 \\
<\mathrm{AM}(X_0,\cL), & \text{for }b<0
\end{array}
\right.
\end{align*}
and
\begin{equation*}
\mathrm{RM}(aX_0+b,a\cL+b)=\mathrm{RM}(X_0,\cL).
\end{equation*}
\end{proposition}
\proof{The proof is trivial once we observe that for $a>0$ and $b\in\mR$
\begin{equation*}
    \infrho(a\cL+b)=a\infrho(\cL)-b,\quad \suprho(a\cL+b)=a\suprho(\cL)-b
\end{equation*}
and $\rho(aX_0+b)=a\rho(X_0)-b$.}

For given $\mu\in\mathbb{R}$ and $\sigma>0$, consider the set
\begin{equation*}
\cL_{\mu,\sigma}=\{X\,:\,\mathbb{E}[X]=\mu,\;\sigma(X)=\sigma\}
\end{equation*}
where the first two moments are fixed. The \emph{standardized version} of $X\in
\cL_{\mu,\sigma}$ is defined by
\begin{equation*}
\widetilde{X}=\frac{X-\mu}{\sigma}\in \cL_{0,1}.
\end{equation*}
Setting $a=1/\sigma$ and $b=-\mu/\sigma$ in Proposition \ref{properties-1}
we immediately obtain
\begin{corollary}
If $\cL\subseteq \cL_{\mu,\sigma}$ and $X_0\in \cL$,
then
\begin{equation*}
\mathrm{RM}(X_0,\cL)=\mathrm{RM}(\widetilde{X}_0,\widetilde{\mathcal{L}}),
\end{equation*}
where $\widetilde{\cL}=\{\widetilde{X}\,:\,X\in \cL\}$. In
particular
\begin{equation*}
\mathrm{RM}(X_0,\cL_{\mu,\sigma})=\mathrm{RM}(\widetilde{X}_0,\cL_{0,1}).
\end{equation*}
\end{corollary}
In what follows we shall be mainly interested in measuring model risk with respect to
$\cL_{\mu,\sigma}$, or some subsets. In view of the last result, we
will concentrate on the particular case $\cL_{0,1}$, provided we
standardize the reference r.v. $X_0$.

Next, we observe that, for fixed $\rho$ and $\cL$, the relative
measure of model risk comes in the form
\begin{equation}  \label{RMR-eq1}
\mathrm{RM}(X_0)=c_1-c_2\rho(X_0),
\end{equation}
where $c_2$ is positive. If $\rho$ is a convex map, as is the case with the
Expected Shortfall, or more generally with the class of (law-invariant)
convex risk measures, then $\mathrm{RM}$ is concave.\footnote{Provided, of course, a certain convex combination of two r.v. in $\cL$
remains in $\cL$.} So, for instance, if $X_1$, $X_2$ and $(X_1+X_2)/2
$ are in $\cL$ and $\mathrm{RM}(X_1)=\mathrm{RM}(X_2)$, then
\begin{equation*}
\mathrm{RM}\left(\frac{X_1+X_2}{2}\right)\geq \frac{\mathrm{RM}(X_1)+\mathrm{RM}(X_2)}{2}=\mathrm{RM}(X_1).
\end{equation*}
Such an inequality can be partly explained by the fact that the model risk
associated with $(X_1+X_2)/2$ is due both to the model risk of the
\emph{marginals} and to the model risk of the joint distribution.

Thanks to \eqref{RMR-eq1}, we see that other possible properties for $\mathrm{RM}$ (like monotonicity, continuity, etc.) are inherited from
similar properties of the risk measure. Subadditivity, a property which is
fulfilled by all coherent risk measures, is an exception. Indeed, if we know
that $\rho(X_1+X_2)\leq \rho(X_1)+\rho(X_2)$, and that $X_1,X_2,X_1+X_2\in
\cL$ we can only conclude that
\begin{equation*}
\mathrm{RM}(X_1+X_2)\geq\mathrm{RM}(X_1)+\mathrm{RM}(X_2)-
\frac{\overline{\rho}(\cL)}{\overline{\rho}(\cL)-\underline{\rho}(\cL)}.
\end{equation*}
and subadditivity is ensured only if the last term in the right hand side is
sufficiently small.
\section{Some examples}
In this section, we illustrate both measures of model risk and study the
following example: we consider a r.v. $X_0$ with a reference
distribution in the set $\cL_{\mu,\sigma}$, which corresponds to the
set of all r.v. with mean $\mu$ and standard deviation $\sigma$,
and we estimate both measures of model risk for two measures of risk, namely
VaR and Expected Shortfall. Without any loss of generality, as previously discussed, we can restrict our
attention to the particular case where the set of r.v. is $\cL_{0,1}$.

Before focusing on our examples, we give a preliminary result on extremal quantiles on a general set $\cL$ that will be useful for the rest of the paper.
\subsection{Preliminary result on extremal quantiles}
Let $\cL$ be a general set of r.v. and $\overline{F}_{\cL}$ and $\underline{F}_{\cL}$ the {\it extremal} functions on $\cL$ defined, for any $x$, as:
\begin{equation*}
 \overline{F}_{\cL}(x)=\sup_{X\in\cL}F_X(x)\qquad \underline{F}_{\cL}(x)=\inf_{X\in\cL}F_X(x).
\end{equation*}
Note that $\overline{F}_{\cL}(+\infty)=1$, $\underline{F}_{\cL}(-\infty)=0$ and that both $\overline{F}_{\cL}$ and $\underline{F}_{\cL}$ are non-decreasing functions\footnote{Note that both $\overline{F}_{\cL}$ and $\underline{F}_{\cL}$ are not necessarily c\`{a}dl\`{a}g. However the set of points on which they are not c\`{a}dl\`{a}g is at most countable}. We will refer to them as the {\it maximal function} and the {\it minimal function} respectively. Note also that these functions are not necessarily distribution functions as it may happen that $\overline{F}_{\cL}(-\infty)>0$ and/or $\underline{F}_{\cL}(+\infty)<1$.
\begin{remark}
If $\overline{F}_{\cL}$ and $\underline{F}_{\cL}$ are indeed distribution functions, they are extremal in the sense of the first order stochastic dominance (denoted $\succcurlyeq_{1sd}$). This means that
\begin{equation*}
    \overline{F}_{\cL}\succcurlyeq_{1sd}F_X \succcurlyeq_{1sd} \underline{F}_{\cL}\qquad \forall X\in\cL
\end{equation*}
and that if $G$ and $H$ are two distribution functions satisfying $G\succcurlyeq_{1sd}F_X \succcurlyeq_{1sd} H$, $\forall X\in\cL$,
then $G\succcurlyeq_{1sd} \overline{F}_{\cL}$ and $\underline{F}_{\cL}\succcurlyeq_{1sd}H$.
\end{remark}
The following result on extremal quantiles will be very useful in the rest of the paper.
\begin{lemma}\label{Lemma extremal functions}
Assume that $\overline{F}_{\cL}(-\infty)<\underline{F}_{\cL}(+\infty)$. If $\overline{F}_{\cL}$ and $\underline{F}_{\cL}$ are invertible functions,\footnote{Except, respectively, on the sets $\{x\,:\,\overline{F}_{\cL}(x)=\overline{F}_{\cL}(-\infty)\text{ or }1\}$ and $\{x\,:\,\underline{F}_{\cL}(x)=0\text{ or }\underline{F}_{\cL}(+\infty)\}$.} then for any $\alpha\in(\overline{F}_{\cL}(-\infty),\underline{F}_{\cL}(+\infty))$ it holds
\begin{equation}\label{lemma-eq}
    \inf_{X\in\cL}q_\alpha(X)=\overline{F}_{\cL}^{-1}(\alpha)\quad\text{ and }\quad \sup_{X\in\cL}q_\alpha(X)=\underline{F}_{\cL}^{-1}(\alpha).
\end{equation}
 If both $\overline{F}_{\cL}$ and $\underline{F}_{\cL}$ are distribution functions, then \eqref{lemma-eq} holds true for any $\alpha \in (0,1)$.
\end{lemma}
\proof{We prove the result for the infimum only, as a similar argument leads to the result for the supremum. If $\alpha>\overline{F}_{\cL}(-\infty)$, then by assumption $a=\overline{F}_{\cL}^{-1}(\alpha)$ is well defined.\\
Let us assume by contradiction that $b=\inf_{X \in \cL} q_{\alpha}(X)>a$. Then for any $X\in\cL$ we have $q_\alpha(X)\geq b>a$, hence $F_X(x)<\alpha$ for $x\in[a,b)$, by the very definition of quantile. It follows that $\overline{F}_{\cL}(x)\leq \alpha=\overline{F}_{\cL}(a)$ for $x\in[a,b)$, but this is in contrast with the fact that, by assumption,  $\overline{F}_{\cL}$ is strictly increasing.\\
If instead we assume that $b<a$, then there exists some $X\in\cL$ such that $q_\alpha(X)<a$. As $\overline{F}_{\cL}$ is strictly increasing, we have
\begin{equation*}
F_X(q_\alpha(X))\leq  \overline{F}_{\cL}(q_\alpha(X))<\overline{F}_{\cL}(a)=\alpha.
\end{equation*}
However, by definition of quantile, it always holds $F_X(q_\alpha(X))\geq \alpha$ and we have reached a contradiction. We then conclude that $b=a$.}

\begin{remark}
The following example underlines the importance of the invertibility of $\overline{F}_{\cL}$ and $\underline{F}_{\cL}$ in Lemma \ref{Lemma extremal functions}. Without this assumption the equalities in \eqref{lemma-eq} need not hold even if we replace $\overline{F}_{\cL}^{-1}$ or $\underline{F}_{\cL}^{-1}$ by the generalized inverses (i.e. the quantile functions). Fix $\alpha$ and consider the sequence $\cL=(X_n)_{n\geq 1}$ of r.v. where $X_n$ takes the value $1$ with probability $1-\alpha+\frac{1}{n}$ and the value $0$ with probability $\alpha -\frac{1}{n}$. It is easy to check that
\begin{equation*}
\overline{F}_{\cL}(x)\equiv \sup_n F_{X_n}(x)=\begin{cases}
               {\displaystyle 0} & \text{if $x< 0$} \\
               \alpha & \text{if $0 \leq x <1$}\\
               1 & \text{if $x \geq 1$}
             \end{cases}
             \end{equation*}
 If $\overline{X}\sim \overline{F}_{\cL}$, we have $q_\alpha(\overline{X})=0$ even though $q_{\alpha}(X_n)=1$ for any $n\geq 1$. So, \eqref{lemma-eq} does not hold in this case.
\end{remark}
\subsection{Model risk for VaR}
Following Section 4 in Royden (1953) and Chapter 3, Section 4 in H\"{u}rlimann (2008), using classical Chebyshev-Markov inequalities, the extremal functions on $\cL_{0,1}$ are distributions and are given as follows:
\begin{equation*}
    \overline{F}_{\cL_{0,1}}(x)=\begin{cases}
               {\displaystyle \frac{1}{1+x^2}} & \text{if $x\leq 0$} \\
               1 & \text{if $x\geq 0$}
             \end{cases}
             \qquad \text{and} \qquad \underline{F}_{\cL_{0,1}}(x)=\begin{cases}
                   0 & \text{if $x\leq 0$} \\
                  {\displaystyle \frac{x^2}{1+x^2}} & \text{if $x\geq 0$}.
             \end{cases}
\end{equation*}
These extremal distributions are often called, respectively, maximal and minimal Chebyshev-Markov distributions for $\cL_{0,1}$. Note, however, that both extremal distributions $\underline{F}_{\cL_{0,1}}$ and $\overline{F}_{\cL_{0,1}}$ are not in $\cL_{0,1}$. In fact, the mean of $\underline{F}_{\cL_{0,1}}$ is negative, the mean of $\overline{F}_{\cL_{0,1}}$ is positive and both variances are infinite.

From Lemma \ref{Lemma extremal functions}, as both $\underline{F}_{\cL_{0,1}}$ and $\overline{F}_{\cL_{0,1}}$ are invertible, the following identities prevail for the extremum quantiles (see for instance H\"{u}rlimann 2002, Theorem 3.1, or Bertsimas et al. 2004, Theorem 2):
\begin{align*}
    &\inf_{X\in \cL_{0,1}}q_{\alpha}(X)=\overline{F}_{\cL_{0,1}}^{-1}(\alpha)=-\sqrt{\frac{1-\alpha}{\alpha}}\\
    &\sup_{X\in \cL_{0,1}}q_{\alpha}(X)=\underline{F}_{\cL_{0,1}}^{-1}(\alpha)=\sqrt{\frac{\alpha}{1-\alpha}}.
\end{align*}
As a straightforward consequence of the extremal quantiles, the following result holds true:
\begin{proposition}
$(i)$ The absolute measure of model risk for $\VAR_{\alpha}$ at $X_0$ is:
$$\AMR(X_0,\cL_{0,1}) = \frac{\sqrt{\frac{1-\alpha}{\alpha}}}{\VAR_{\alpha}(X_0)}-1.$$
$(ii)$ The relative measure of model risk for $\VAR_{\alpha}$ at $X_0$ is:
$$\RMR(X_0,\cL_{0,1}) = \frac{\sqrt{\frac{1-\alpha}{\alpha}}-\VAR_{\alpha}(X_0)}{\sqrt{\frac{1-\alpha}{\alpha}}+\sqrt{\frac{\alpha}{1-\alpha}}}=
(1-\alpha)-\sqrt{\alpha(1-\alpha)}\VAR_{\alpha}(X_0).$$
\end{proposition}
This result will be illustrated later in Subsection \ref{Sunsection Illustration}.
\begin{remark}
Note that $\sup_{X
\in \cL_{0,1}}\VAR_{\alpha}(X)>0$ and $\inf_{X\in \cL_{0,1}}\VAR_{\alpha}(X)<0$. Therefore, in the class
$\cL_{0,1}$, some distributions are acceptable, meaning that they have negative risk, while
others are not. In the case of $\cL_{\mu,\sigma}$, when $\mu >0$, if
$\alpha> \frac{\sigma^2}{\mu^2+\sigma^2}$, then all distributions are
acceptable. When $\mu<0$, if $\alpha <\frac{\mu^2}{\mu^2+\sigma^2}$, then all distributions are
non-acceptable.
\end{remark}
\begin{remark}
As pointed out by H\"{u}rlimann (2008) (Chapter 4,
Section 3), knowledge of the skewness does not improve the
Chebyshev extremal distributions when considering distributions over
$(-\infty,+\infty)$. Therefore, if $X_0\in\cL_{\mu,\sigma}$:
\begin{align*}
    &\AMR(X_0,\cL_{\mu,\sigma,\xi})=\AMR(X_0,\cL_{\mu,\sigma})\\
    &\RMR(X_0,\cL_{\mu,\sigma,\xi})=\RMR(X_0,\cL_{\mu,\sigma}).
\end{align*}
where $\cL_{\mu,\sigma,\xi}=\{X\in\cL_{\mu,\sigma}\,:\,\xi(X)=\xi(X_0)\}$ and $\xi(X)$ denotes the skewness of $X$.
\end{remark}
\subsection{Model risk for Expected Shortfall}
Adopting a similar approach for the Expected Shortfall is not so easy since the Lemma \ref{Lemma extremal functions} gives a result on the extremal quantiles, but not on the extremal Expected Shortfalls. However, a recent result by Bertsimas et al. (2004) (Theorem 2) using arguments from convex analysis gives the following identities for the extremal Expected Shortfalls on the set $\cL_{0,1}$:
\begin{align}\label{Extremal ES}
&\inf_{X\in \cL_{0,1}}\mathrm{ES}_{\alpha}(X)=0 \\
&\sup_{X\in \cL_{0,1}}\mathrm{ES}_{\alpha}(X)=\sqrt{\frac{1-\alpha}{\alpha}}.
\end{align}
To our knowledge, similar results for a general set $\mathcal{L}$ have not been obtained.

As a straightforward consequence, the following result on model risk holds true:
\begin{proposition}
$(i)$ The absolute measure of model risk for $\mathrm{ES}_{\alpha}$ at $X_0$
is:
\begin{equation*}
\mathrm{AM}(X_0,\cL_{0,1})= \frac{\sqrt{\frac{1-\alpha}{\alpha}}}{\mathrm{ES}_{\alpha}(X_0)}-1.
\end{equation*}
$(ii)$ The relative measure of model risk for $\mathrm{ES}_\alpha$ at $X_0$ is:
\begin{equation*}
\mathrm{RM}(X_0,\cL_{0,1}) = \frac{\sqrt{\frac{1-\alpha}{\alpha}}-\mathrm{ES}_{\alpha}(X_0)}{\sqrt{\frac{1-\alpha}{\alpha}}}= 1-\sqrt{\frac{\alpha}{1-\alpha}}\mathrm{ES}_{\alpha}(X_0).
\end{equation*}
\end{proposition}
This result will be illustrated later in Subsection \ref{Sunsection Illustration}.
\begin{remark}
As mentioned earlier, we cannot use Lemma \ref{Lemma extremal functions} to obtain the Extremal Shortfalls. However, we may wonder whether the Extremal Shortfalls in \eqref{Extremal ES} are obtained as Expected Shortfalls of some extremal distributions. Since the Expected Shortfall is monotone with respect to the stop-loss order (see for instance B\"{a}uerle and M\"{u}ller (2006)), we look at the extremal distributions for the stop-loss order on the set $\cL_{0,1}$. Following H\"{u}rlimann (2002), we use the fact that the stop-loss transform for a distribution $F$ is defined as:
\begin{equation*}
  \Pi_F(x) = \int_x^{\infty} (1-F(y))dy.
\end{equation*}
By simple calculation, we have:
\begin{equation*}
    F(x)=1+\Pi'_F(x).
\end{equation*}
Such a relationship also holds true for the extremal stop-loss distributions (see for instance Equation (1.3) in H\"{u}rlimann (2002)):
\begin{equation*}
  F^{SL}_{\max}(x)=1+\Pi_{\max}'(x),
\end{equation*}
where
\begin{equation*}
  \Pi_{\max}(x) \equiv  \sup_{F \in \cL_{0,1}}\Pi_F(x),
\end{equation*}
and the same holds true for the infimum.

Therefore, in order to get the extremal stop-loss distributions, we first need to obtain the extremal stop-loss transforms. For the maximum stop-loss transform, we refer to Theorem 2 in Jansen et al. (1986) and obtain:
\begin{equation*}
    \Pi_{\max}(x) = \frac{\sqrt{x^2+1}-x}{2}.
\end{equation*}
For the minimum stop-loss transform, we refer to Table 5.2 Section 5., Chapter 3 in H\"{u}rlimann (2008):
\begin{equation*}
    \Pi_{\min}(x)=\begin{cases}
                   -x & \text{if $x\leq 0$} \\
                   0 & \text{if $x\geq 0$}.
             \end{cases}
\end{equation*}
Finally, we obtain the extremal stop-loss distributions:
\begin{equation*}
    F^{SL}_{\max}(x)=\frac{1}{2}\left(1+\frac{x}{\sqrt{x^2+1}}\right)
             \qquad \text{and} \qquad F^{SL}_{\min}(x)=\begin{cases}
                   1 & \text{if $x\geq 0$} \\
                   0 & \text{if $x< 0$}.
             \end{cases}
\end{equation*}
We finally obtain, using Equation \eqref{Extremal ES} that:
$$\ES_{\alpha}(F^{SL}_{\min})=0=\inf_{X
\in\cL_{0,1}}\ES_{\alpha}(X)$$
and
$$\ES_{\alpha}(F^{SL}_{\max})=\sqrt{\frac{1-\alpha}{\alpha}}=
\sup_{X\in\cL_{0,1}}\ES_{\alpha}(X).$$
Note that using the extremal distributions $\underline{F}_{\cL_{0,1}}$ and $\overline{F}_{\cL_{0,1}}$ for the first-order stochastic dominance will give us some bounds which are not sharp as discussed earlier in Subsection \ref{Subsection sharp bounds} (in particular Equation \eqref{ES-cantelli}).
\end{remark}
\subsection{Illustration}\label{Sunsection Illustration}
\begin{figure}[htbp]
 \centering
 \subfigure[$\VAR_\alpha$]
   {\includegraphics[width=7.8cm,height=6.5cm]{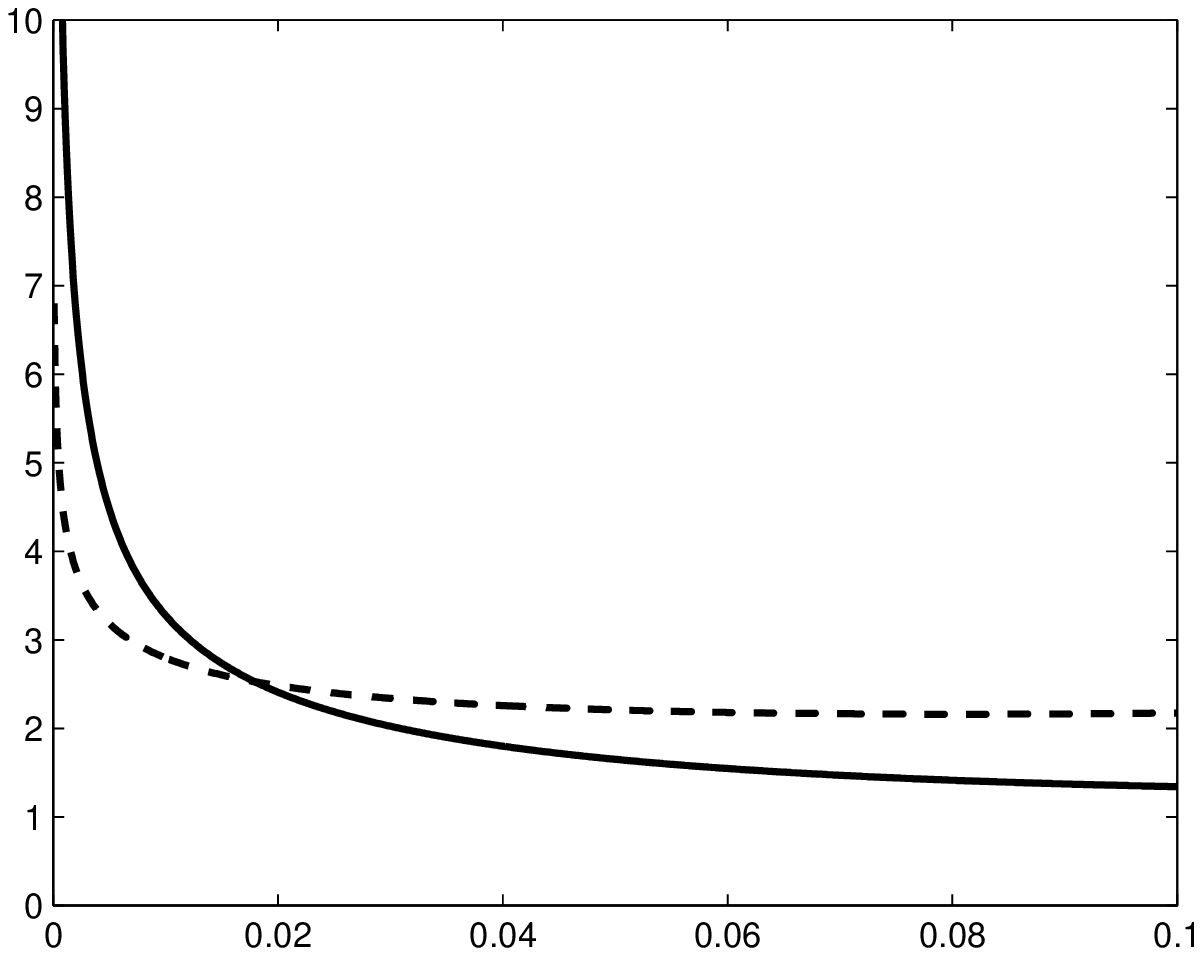}}
 \subfigure[$\ES_\alpha$]
   {\includegraphics[width=7.8cm,height=6.5cm]{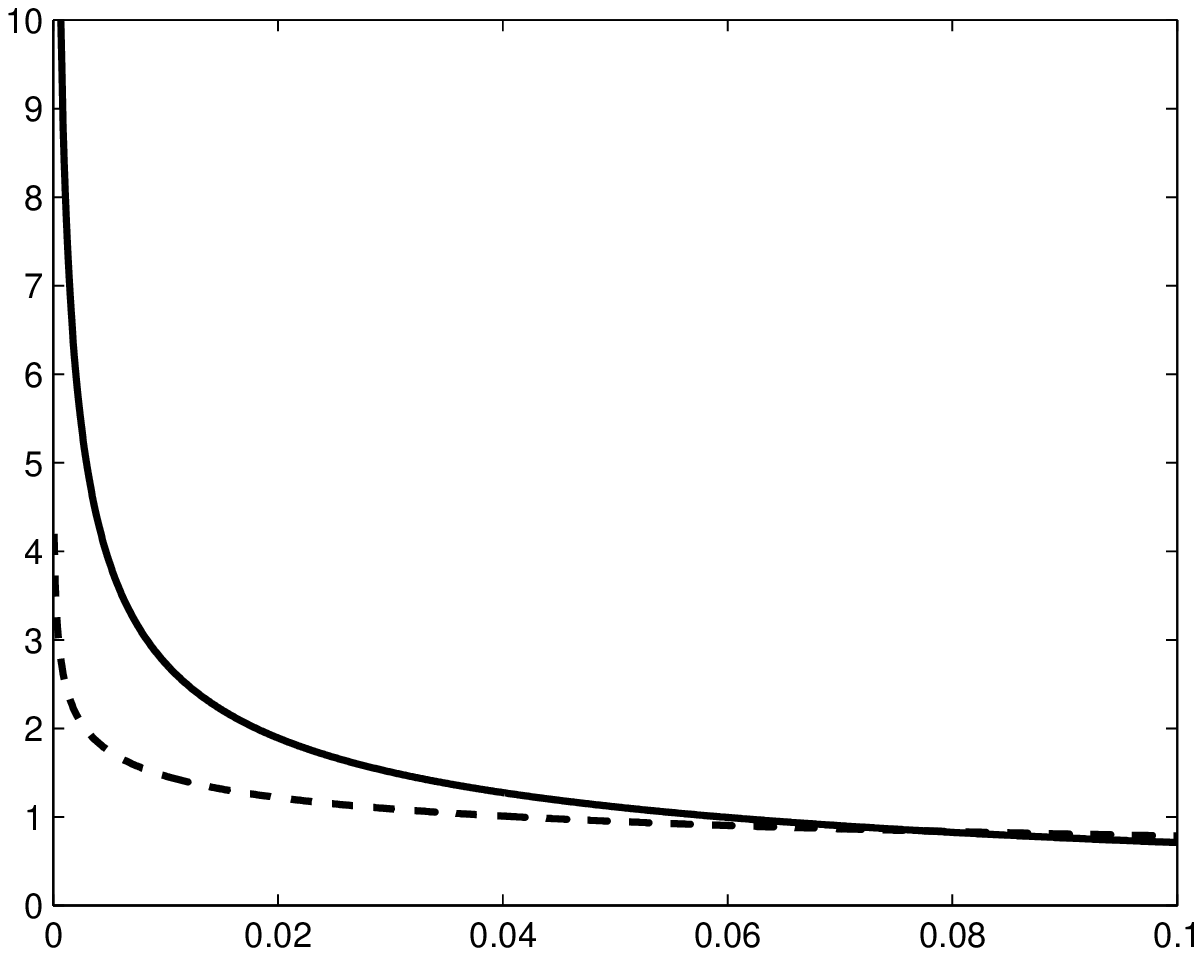}}
 \caption{Absolute measure of model risk as a function of $\alpha$. Continuous lines: $X_0$ standard normal. Dashed lines: $X_0$ Student-t with $\nu=3$ degrees of freedom. \label{figD}}
 \end{figure}

\begin{figure}[h]
 \centering
 \subfigure[$\VAR_\alpha$]
   {\includegraphics[width=7.8cm,height=6.5cm]{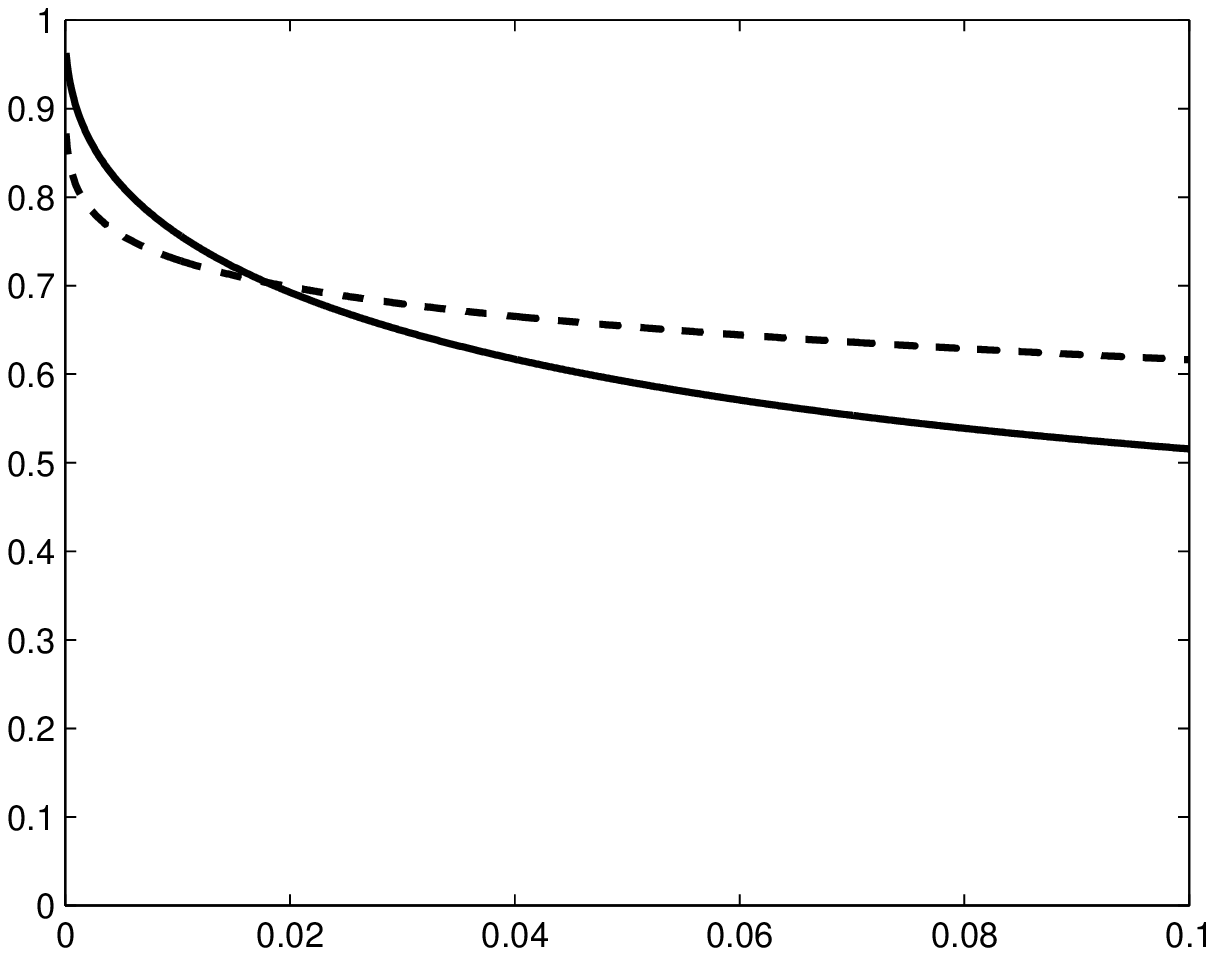}}
 \subfigure[$\ES_\alpha$]
   {\includegraphics[width=7.8cm,height=6.5cm]{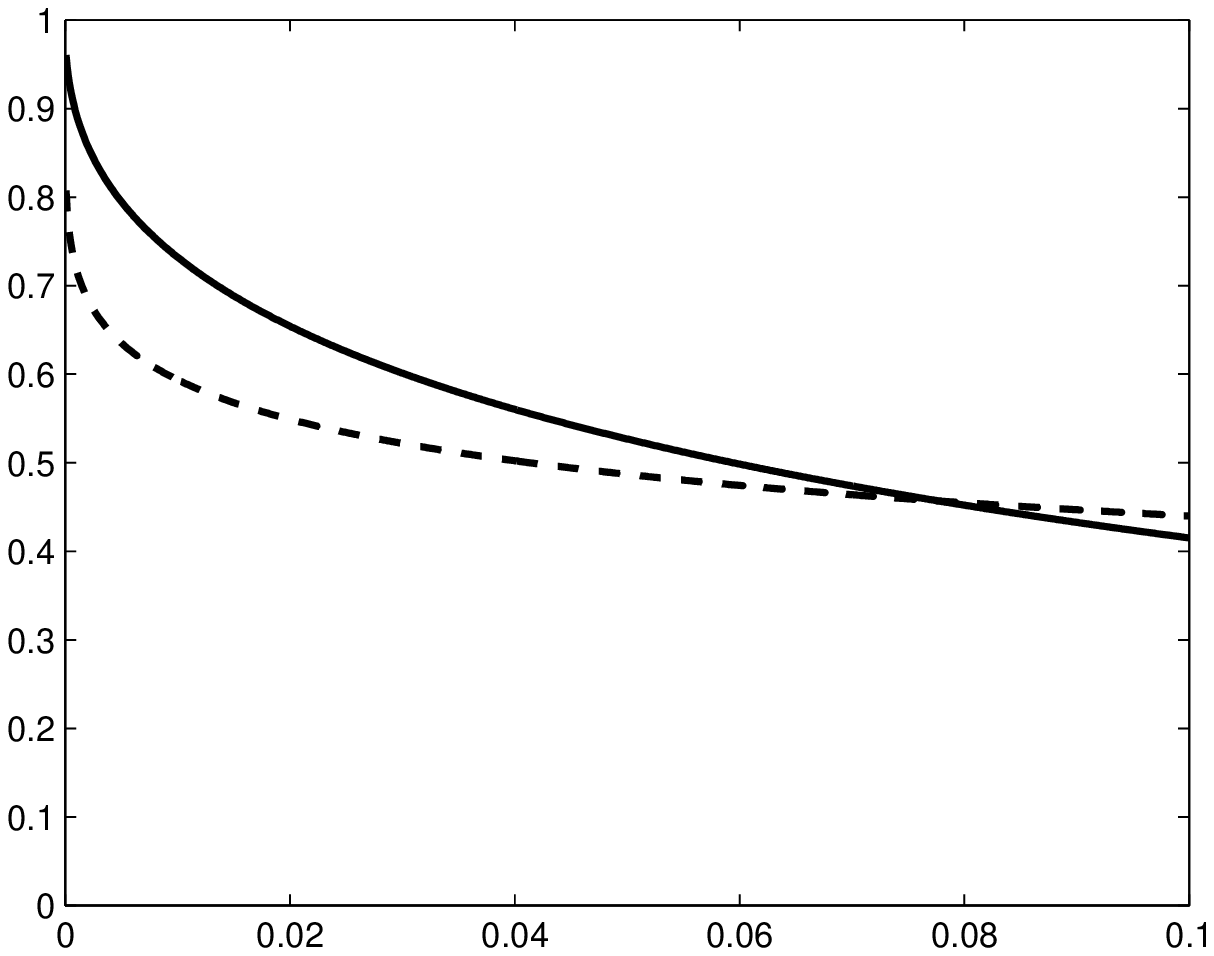}}
 \caption{Relative measure of model risk as a function of $\alpha$. Continuous lines: $X_0$ standard normal. Dashed lines: $X_0$ Student-t with $\nu=3$ degrees of freedom. \label{figDD}}
 \end{figure}

We numerically compute both measures of model risk for
standard (i.e. in $\cL_{0,1}$) r.v. following the normal or
Student-t distribution. We are especially interested in the dependence of the
measures on the order $\alpha$ of the Value at Risk or the Expected
Shortfall. This dependence is depicted in Figures \ref{figD} and \ref{figDD}.

 It is natural to expect that using a reference fat-tailed distribution (Student-t) yields lower model risk than starting with a normal one. While for the Expected Shortfall this is true for any practical\footnote{Precisely, for $\alpha\lessapprox 8\%$. See Figures \ref{figD} (right) and \ref{figDD} (right).} value of $\alpha$, for the Value-at-Risk this holds only for $\alpha$ small enough ($\alpha\lessapprox 1.5\%$).

 We can also notice that the relative measure of model risk, for both VaR and Expected Shortfall and for both distributions, goes to $1$ as $\alpha\to 0$. In other words, as we go further in the (left) tails, any given distribution departs more and more from the worst case. We think this is a general behaviour, although we offer no proof for this claim.

\begin{figure}[htbp]
 \centering
 \subfigure[Absolute measure]
   {\includegraphics[width=7.8cm,height=6.5cm]{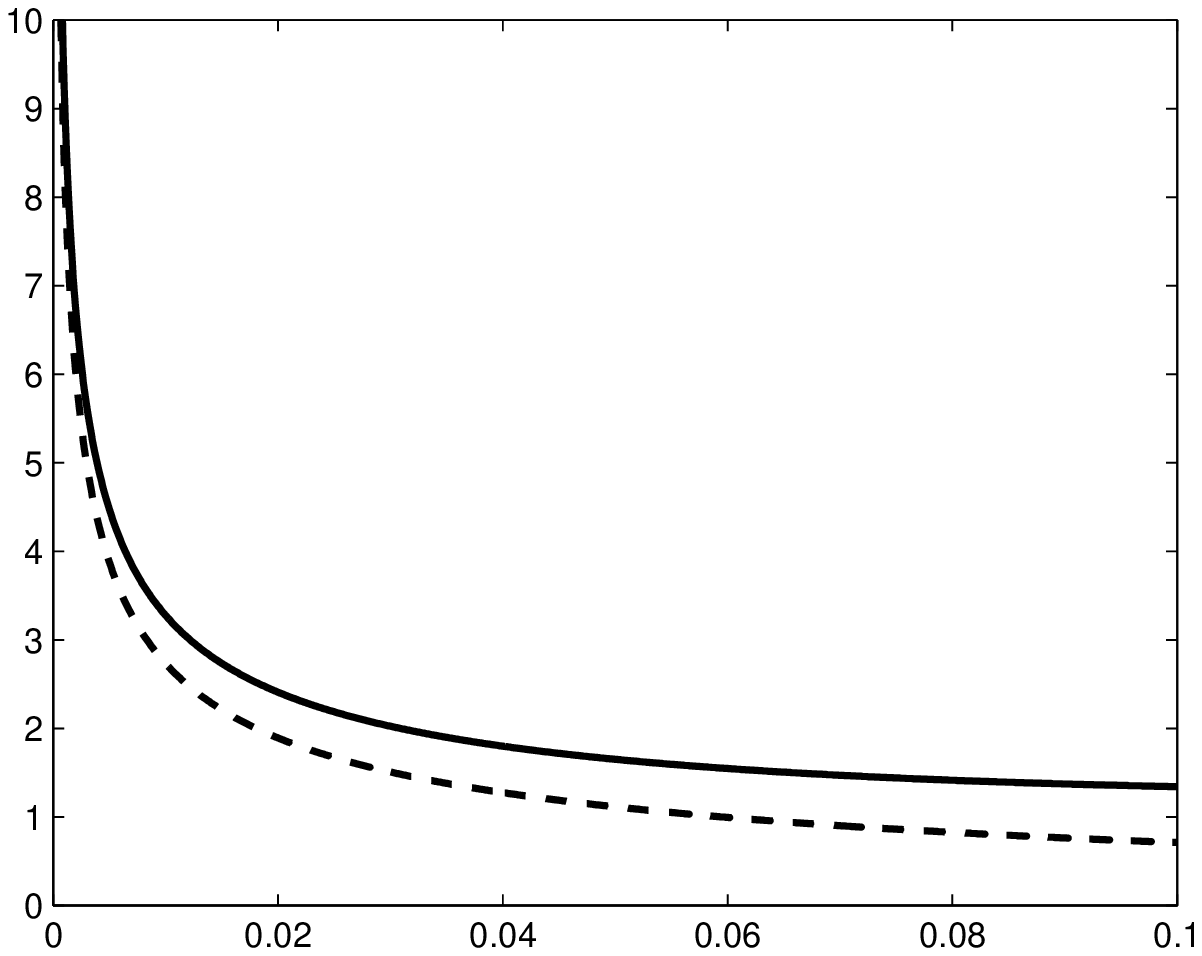}}
 \subfigure[Relative measure]
   {\includegraphics[width=7.8cm,height=6.5cm]{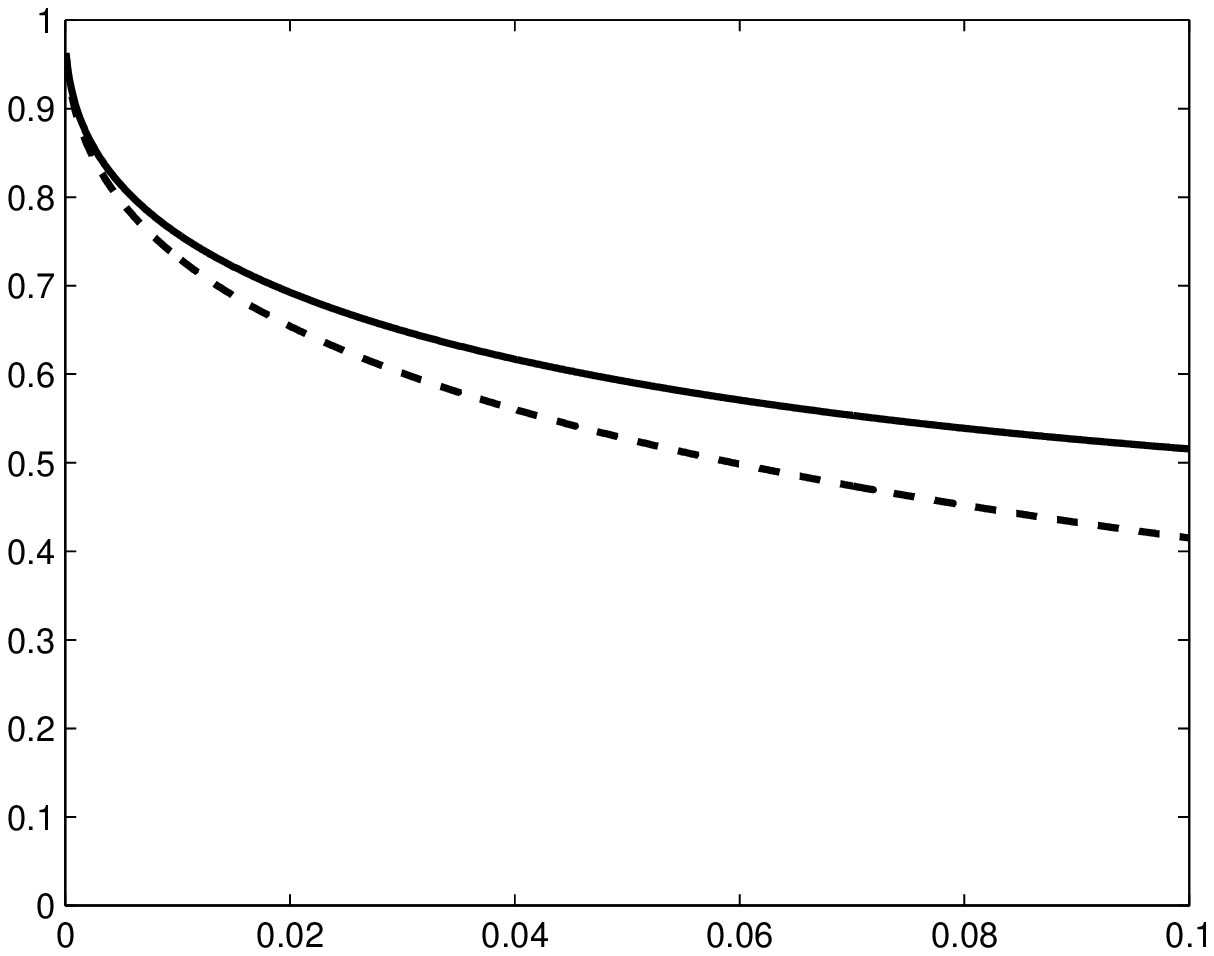}}
 \caption{Absolute and relative measure of model risk as a function of $\alpha$ with $X_0$ standard normal. Continuous lines: VaR. Dashed lines: Expected Shortfall. \label{figDDD}}
 \end{figure}

 The graphs in Figure \ref{figDDD} compare the absolute (left) and relative (right) measure of model risk for VaR and Expected Shortfall, using a normal reference distribution. We see that in both cases the Expected Shortfall has a lower level of model risk. By taking a Student-t as the reference distribution we obtain a similar behaviour. This is probably at odds with what we would expect: indeed, it is often said that Expected Shortfall is more sensitive to the model choice than VaR as the former depends on the whole left tail.\footnote{See also the related discussion on statistical robustness in Cont et al. (2010).} Instead, at least with respect to our two measures of model risk, the opposite proves true.

\section{Local measure of model risk}
In this section we introduce a \emph{local} measure of model risk, by taking the limit of the
relative measure $\mathrm{RM}$ on a family of perturbation sets that shrink
to the singleton $\{X_0\}$. This measure attempts to assess model risk for infinitesimal perturbations.
\subsection{The definition}
Let $(\cL_{\eps})_{\eps> 0}$
be a family of sets, each one contained in $\cL_\rho$ and such that
\begin{equation*}
\cL_{\eps}\searrow \{X_0\}\quad\text{ as }\eps\to 0.
\end{equation*}
This means that $\cL_{\eps}\subset \cL%
_{\eps^{\prime }}$ whenever $\eps< \eps'$ and $\cap_{\eps>0} \cL_{\eps}=\{X_0\}$. Below, we will see
some examples based on distances and on mixtures.
\begin{definition}
The \textbf{local measure of model risk} associated to $\rho$, $X_0$ and the
family $(\cL_{\eps})_{\eps> 0}$ is
\begin{equation*}
\mathrm{LM}=\lim_{\eps\to 0}\mathrm{RM}(X_0,\cL_\eps)=\lim_{\eps\to 0}\frac{\overline{\rho}(\cL_\eps)-\rho(X_0)}{\overline{\rho}(\cL_\eps)-\underline{\rho}(\cL_\eps)},
\end{equation*}
provided the limit exists.
\end{definition}
The limit defining $\mathrm{LM}$ is evidently in the form $0/0$; however, if
it exists, then it is in the interval $[0,1]$ as $\mathrm{RM}(X_0,L_\eps)\in[0,1]$
for any $\eps$. The local measure
describes the \textit{relative position} of $\rho(X_0)$ with respect to  the
worst and best cases for infinitesimal perturbations.
\subsection{An example based on distances}
In what follows, we will consider the case $\rho=\mathrm{VaR}_\alpha$ for some $\alpha$, so
that $\cL_\rho$ is the set of all r.v. and we will make no reference
at it in the definition of $\cL_\eps$. As a first example of computation of the local model risk, consider
the family of sets defined by:
\begin{equation}  \label{set-distance}
\cL_\eps=\{X\,:\, d(X,X_0)\leq \eps \},
\end{equation}
where $d$ is some given distance between distributions. It can immediately be recognized that such a family satisfies the assumptions stated above. In
particular, we can consider the Kolmogorov (or uniform) distance
\begin{equation*}
d_K(X,Y)=\sup_{x\in\mathbb{R}}|F_X(x)-F_Y(x)|
\end{equation*}
or the L\'{e}vy distance
\begin{equation*}
d_L(X,Y)=\inf\{a>0\,:\,F_X(x-a)-a\leq F_Y(x)\leq F_X(x+a)+a\;\forall x\in\mathbb{R}\}.
\end{equation*}
\begin{proposition}
If $\rho=\mathrm{VaR}_\alpha$ for $\alpha\in(0,1)$ and the family $(\mathcal{L}_\eps)$ is defined as in \eqref{set-distance}, with $d=d_K$ or $d=d_L$,
then
\begin{equation*}
LM(X_0,(\cL_\eps))=\frac{1}{2}
\end{equation*}
for any absolutely continuous r.v. $X_0$.
\end{proposition}
\proof{If $d=d_K$ it can immediately be seen that
\begin{equation}  \label{maximalF-distance}
\overline{F}_{\cL_\eps}(x)=\min\{F_0(x)+\eps,1\},\quad
\underline{F}_{\cL_\eps}(x)=\max\{F_0(x)-\eps,0\}.
\end{equation}
From now on, let $\eps<\min\{\alpha,1-\alpha\}$, so that
\begin{equation*}
    \overline{F}_{\cL_\eps}(-\infty)=\eps<\alpha<1-\eps=\underline{F}_{\cL_\eps}(+\infty).
\end{equation*}
By assumption, $F_0$ is invertible and therefore both $\overline{F}_{\cL_\eps}$ and $\underline{F}_{\cL_\eps}$ are invertible; an immediate computation shows that
\begin{equation*}
    \overline{F}_{\cL_\eps}^{-1}(\alpha)=F_0^{-1}(\alpha-\eps),\quad \underline{F}_{\cL_\eps}^{-1}(\alpha)=F_0^{-1}(\alpha+\eps).
\end{equation*}
We can then apply Lemma \ref{Lemma extremal functions}, obtaining
\begin{equation*}
\sup_{X\in\cL_\eps}\mathrm{VaR}_\alpha(X)=-F_0^{-1}(\alpha-\eps),\quad
\inf_{X\in\cL_\eps}\mathrm{VaR}_\alpha(X)=-F_0^{-1}(\alpha+\eps)
\end{equation*}
and therefore
\begin{equation*}
\mathrm{LM}=\lim_{\eps\to 0}\frac{F_0^{-1}(\alpha)-F_0^{-1}(\alpha-\eps)} {F_0^{-1}(\alpha+\eps)-F_0^{-1}(\alpha-\eps)},
\end{equation*}
as $\VAR_\alpha(X_0)=-F_0^{-1}(\alpha)$. Finally, if $f_0=F^{\prime }_0$ is the density of $X_0$, by applying de l'H\^{o}pital's
rule we have
\begin{equation*}
\mathrm{LM}=\lim_{\eps\to 0}\frac{1/f_0(\alpha-\eps)}{1/f_0(\alpha+\eps)+1/f_0(\alpha-\eps)}=\frac{1}{2}
\end{equation*}
In the case $d=d_L$ we start by observing that $\overline{F}_{\cL_\eps}(x)=
\min\{F_0(x+\eps)+\eps,0\}$ and $\underline{F}_{\cL_\eps}(x)=\max\{F_0(x-\eps)-\eps,0\}$ and then proceed similarly as above.}

This result is quite natural as the set of perturbations is in a sense
asymptotically symmetrical around $X_0$. Therefore the relative measure of
model risk converges to $1/2$. However, we stress that this is true only in
the limit $\eps\to 0$ and not for a fixed $\eps$.
\subsection{An example based on mixtures}
Let $F_0$ be the distribution of $X_0\in\cL_{0,1}$; for $\eps<1$ define
\begin{equation}  \label{mixtures}
\cL_\eps=\{X\,:\, X\sim (1-\theta)F_0+\theta F_Y,\;Y\in
\cL_{0,1},\;\theta\in[0,\eps]\}.
\end{equation}
The set $\cL_\eps$ collects all (r.v. distributed as) mixtures between $F_0$ and a
distribution of a standard r.v. $Y$, for which the alternative distribution ($F_Y$) is not weighted too much. It is worth noting that $\cL%
_\eps\subset \cL_{0,1}$ for any $\eps$: indeed, both
the mean and the variance are affine functions of the distributions.
\begin{remark}
We stress that $(1-\theta)F_0+\theta F_Y$ is in general \emph{not} the
distribution of $(1-\theta)X_0+\theta Y$, even if we assume $X_0$ and $Y$
to be independent. Rather, it is the distribution of $(1-I_{A})X_0+I_A Y$, where $A
$ is an event of probability $\theta$, independent from both $X_0$ and $Y$,
and $I_A$ denotes its indicator function.
\end{remark}
\begin{proposition}
If $\rho=\mathrm{VaR}_\alpha$ for $\alpha\in(0,1)$ and the family $(\mathcal{L}_\eps)$ is defined as in \eqref{mixtures}, then
\begin{equation*}
   \mathrm{LM}=1-\alpha(1+\mathrm{VaR}_\alpha(X_0)^2)
\end{equation*}
for any absolutely continuous r.v. $X_0$ for which $\VAR_\alpha(X_0)\geq0$.
\end{proposition}
\proof{The maximal function for $\cL_\eps$ is
\begin{align*}
\overline{F}_{\cL_\eps}(x)&=\sup_{\theta\in
[0,\eps]}\sup_{Y\in\cL_{0,1}}\left\{(1-\theta)F_0(x)+\theta
F_Y(x)\right\} \\
&=\sup_{\theta\in [0,\eps]}\left\{(1-\theta)F_0(x)+\theta \overline{F}_{\cL_{0,1}}(x)\right\} \\
&=(1-\eps)F_0(x)+\eps\overline{F}_{\cL_{0,1}}(x),
\end{align*}
where we have used $\overline{F}_{\cL_{0,1}}(x)-F_0(x)\geq 0$ in
deriving the last equality. Since both $F_0$ and $\overline{F}_{\cL_{0,1}}$ are invertible (the former by assumption), $\overline{F}_{\cL_\eps}$ too is invertible and therefore, applying Lemma \ref{Lemma extremal functions}, we have
\begin{equation}\label{supremum}
    \sup_{X\in\cL_\eps}\VAR_\alpha(X)=-\overline{F}_{\cL_\eps}^{-1}(\alpha).
\end{equation}
Using a similar argument, we find that
\begin{equation}\label{infimum}
    \inf_{X\in\cL_\eps}\VAR_\alpha(X)=-\underline{F}_{\cL_\eps}^{-1}(\alpha),
\end{equation}
where $\underline{F}_{\cL_\eps}(x)=(1-\eps)F_0(x)+\eps\underline{F}_{\cL_{0,1}}(x)$. As a consequence, the local measure of model risk is
\begin{equation}\label{limit}
    LM=\lim_{\eps\to 0}\frac{-\overline{F}_{\cL_\eps}^{-1}(\alpha)-\VAR_\alpha(X_0)}
    {-\overline{F}_{\cL_\eps}^{-1}(\alpha)+\underline{F}_{\cL_\eps}^{-1}(\alpha)}.
\end{equation}
If we set $\overline{\psi}(\eps)=\overline{F}_{\cL_\eps}^{-1}(\alpha)$, then, by definition
\begin{equation*}
    (1-\eps)F_0(\overline{\psi}(\eps))+\eps \overline{F}_{\cL_{0,1}}(\overline{\psi}(\eps))=\alpha.
\end{equation*}
Differentiating (in $\eps$) both sides, we obtain
\begin{equation*}
    f_0(\overline{\psi})\overline{\psi}'+\overline{F}_{\cL_{0,1}}(\overline{\psi})-
    F_0(\overline{\psi})+
    \eps(\overline{F}_{\cL_{0,1}}'(\overline{\psi})-f_0(\overline{\psi})\overline{\psi}'=0,
\end{equation*}
where $f_0=F_0'$ is the density of $X_0$. Setting $\eps=0$ and observing that $\overline{\psi}(0)=F_0^{-1}(\alpha)=-\VAR_\alpha(X_0)$, so that $F_0(\overline{\psi}(0))=\alpha$, we readily obtain\footnote{A similar proof can also be found in Barrieu and Ravanelli (2013)}
\begin{equation*}
    \overline{\psi}'(0)=
    \frac{\alpha-\overline{F}_{\cL_{0,1}}(-\VAR_\alpha(X_0))}{f_0(-\VAR_\alpha(X_0))}.
\end{equation*}
In a very similar way, we can prove that $\underline{\psi}(\eps)=\underline{F}_{\cL_\eps}^{-1}(\alpha)$ satisfies
\begin{equation*}
    \underline{\psi}'(0)=\frac{\alpha-\underline{F}_{\cL_{0,1}}(-\VAR_\alpha(X_0))}{f_0(-\VAR_\alpha(X_0))}.
\end{equation*}
Applying de l'H\^{o}pital's rule to \eqref{limit} and simplifying the result we obtain
\begin{equation*}
    LM=\frac{\overline{F}_{\cL_{0,1}}(-\VAR_\alpha(X_0))-\alpha}
    {\overline{F}_{\cL_{0,1}}(-\VAR_\alpha(X_0))-\underline{F}_{\cL_{0,1}}(-\VAR_\alpha(X_0))}.
\end{equation*}
As $-\VAR_\alpha(X_0)\leq 0$ by assumption, we have
\begin{equation*}
    \overline{F}_{\cL_{0,1}}(-\VAR_\alpha(X_0))=\frac{1}{1+\VAR_\alpha(X_0)^2},\quad \underline{F}_{\cL_{0,1}}(-\VAR_\alpha(X_0))=0
\end{equation*}
and we reach the final result as an immediate computation.}
\begin{remark}
Remembering the form of $\overline{F}_{\cL_{0,1}}$ and $\underline{F}_{\cL_{0,1}}$, from \eqref{supremum} and \eqref{infimum} it easily follows that, if $\alpha$ is not too large,\footnote{It is sufficient to assume $\alpha\leq (1-\eps)F_0(0)$}  $r=\sup_{X\in\cL_{\eps}}\VAR_\alpha(X)$ is the unique solution of
    \begin{equation*}
    (1-\eps)F_0(-r)+\frac{\eps}{1+r^2}=\alpha,
\end{equation*}
while
\begin{equation*}
    \inf_{X\in\cL_{\eps}}\VAR_\alpha(X)=\VAR_{\frac{\alpha}{1-\eps}}(X_0).
\end{equation*}
This result allows us to compute the relative measure of model risk with respect to $\cL_\eps$ for finite values of $\eps$.
\end{remark}
As an illustration we compute the local measure of model risk when $X_0$ is standard normal or Student-t (see Figure 6). Consistent with the observations we made in the last section, regarding the relative measure, we see that starting with a fat-tailed reference distribution yields a lower local measure with respect to a normal distribution only when $\alpha$ is small enough.

\begin{figure}[htbp]
 \centering
 \includegraphics[width=9cm,height=7cm]{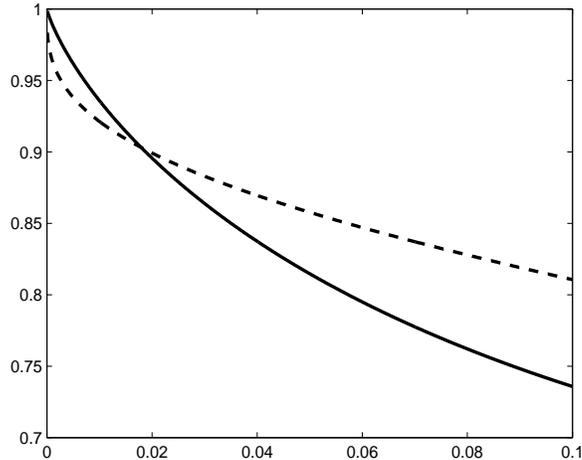}
 \caption{Local measure of model risk for VaR as a function of $\alpha$. Continuous line: $X_0$ standard normal. Dashed line: $X_0$ Student-t with $\nu=3$ degrees of freedom. \label{fig100}}
 \end{figure}
%
\section{Conclusion}
The study of the impact of model risk and its quantification is an essential part of the whole risk measurement procedure. In this paper, we introduce three quantitative measures of the model risk when choosing a particular reference model within a given class: the absolute measure of model risk, the relative measure of model risk and the local measure of model risk. Each of the measures we propose has a specific purpose and so allows for flexibility in their use. We obtain explicit formulae in some interesting cases, in order to emphasize the practicability and tractability of our approach. However, our contribution is not limited to the study of these particular examples and our measures of model risk can be applied to more general settings.


\end{document}